%% file: agujournaltemplate.tex
\documentclass[draft]{agujournal2019}
\usepackage{url} 
\usepackage{lineno}
\usepackage[inline]{trackchanges} 
\usepackage{soul}

\usepackage{graphicx} 
\usepackage{parskip}
\usepackage{derivative}
\usepackage{amsmath}
\usepackage{amssymb}
\usepackage{multicol}
\usepackage{enumitem}
\usepackage{bm}
\usepackage{url}
\draftfalse

\journalname{Earth and Space Science}

\renewcommand{\deg}{^\circ}

\newcommand{\x}{\cdot}

\begin{document}

%
%


\title{Extending Temporal Disturbance Estimations For Magnetic Anomaly Navigation and Mapping}

%
%




\authors{Anutam Srinivasan\affil{1,2}\thanks{This work was completed during an internship at the Autonomy \& Navigation Technology Center. He has now moved to Georgia Institute of Technology.}, Aaron Nielsen\affil{2}}


\affiliation{1}{Ohio State University}
\affiliation{2}{Autonomy \& Navigation Technology Center, Air Force Institute of Technology}




\correspondingauthor{Anutam Srinivasan}{asrinivasan350@gatech.edu}



\begin{keypoints}
\item In practice, measurements of geomagnetic temporal variations from a ground station are considered valid within a 100 km radius.  
\item To extend this radius, we propose the Extended Reference Station Model and evaluate it with radii over several thousand kilometers. 
\item Our data-driven approach has a median root mean square error below 10 nT over land masses and water bodies.
\end{keypoints}

%
%

%
%


\begin{abstract}
\input{src/abstract}
\end{abstract}

\section*{Plain Language Summary}
\input{src/pls}

%
%

%


%
%
%
%

\section{Introduction}
\input{src/introduction}

\section{Preliminaries} \label{sec:prelim}
\input{src/prelim}

\section{Extended Reference Station Model} 
\input{src/ersm_method}

\section{Testing the Model}
\input{src/exp_method}

\section{Results and Analysis} \label{sec:results}
\input{src/results}

\section{Conclusion}
\input{src/conclusion}
\appendix
\section{DIFI Model}
\input{src/appendix}
\section*{Open Research Section}
\textbf{Data:} The data used to support our conclusions was accessed through \texttt{INTERMAGNET} \cite{Intermagnet2021-fd} and can be downloaded from the following link \texttt{\url{https://imag-data.bgs.ac.uk/GIN_V1/GINForms2}}. The $K_p$ data was retrieved from GFZ Potsdam's database \cite{Matzka2021-zn}, and is available at the following link \texttt{\url{https://kp.gfz-potsdam.de/kpdata?startdate=YYYY-MM-DD\&enddate=YYYY-MM-DD\&format=kp2\#kpdatadownload-14}}, where YYYY-MM-DD are placeholders for the start and end dates of the required data. 

\textbf{Software:} The repository and software (Jupyter notebooks) to apply ERSM are available at \texttt{\url{https://github.com/AFIT-EENG-MagNav/ERSM}} (version=1.0.2), along with the Jupyter notebooks to produce figures using results from our model, are available in the same repository \cite{Nielsen_AFITEENGMagNavERSM_2025}. The downloaded \texttt{INTERMAGNET} is also available in this repository. We also provide YAML files to set up the conda environment used to run the models and data visualization.

\acknowledgments
\input{src/acknowledgement}

%
%

\bibliography{apa_citations}

%
%
%
%
%

\end{document}

%% file: src/abstract.tex
Slow-moving vehicles relying on crustal magnetic anomaly navigation (MagNav) or vehicles revisiting the same location in a short time - such as those used for surveys in magnetic anomaly mapping -  require fixed ground stations within 100 km of the vehicle's trajectory to measure and remove the geomagnetic disturbance field from magnetic readings. This approach is impractical due to the limited network of fixed-ground magnetometer stations, making long-range (several hundred kilometers long) aeromagnetic surveys for anomaly map-making infeasible. To address these challenges, we developed the Extended Reference Station Model (ERSM). ERSM applies a longitudinal correction and regression model to an extended reference ground magnetometer station (ERS) to produce an estimate of the local temporal disturbance field. ERSM is regression model-agnostic, so we implemented a linear regression, a k-nearest neighbors (kNN) regression, and a neural-network regression model to assess performance benefits. Our results show typical performance below 10nT root mean square error and median performance below 5nT for typical use with the kNN and neural-net model for farther distances and below 5nT performance using the linear regression model on stations with proximity. We also consider how space-weather events, water-body separation, and proximity to polar regions affect the model performance based on ERS selection. 

%% file: src/pls.tex
Recent field tests have shown that magnetic anomaly navigation (MagNav) is a viable alternative to GPS-based navigation. MagNav relies on successfully measuring static magnetic anomaly fields—i.e., anomalies—caused by the Earth's crust and geographic landscape while removing Earth's core field and dynamic components from measurements. Current standard practice, for MagNav and anomaly mapping, uses a fixed ground station with magnetic field sensors (magnetometers) within 100 km of a vehicle's location to measure and subtract the diurnal variations (DV) – or 1-day periodic variations – of the Earth's magnetic field. However, this is impractical for long-distance travel or surveys far from land. The Extended Reference Station Model (ERSM) discussed here applies a correction to extended ground stations (farther than 100 km) to estimate the DV near a vehicle. We evaluate how longitudinal and latitude displacement between the extended station and the location of the desired prediction affects the ERSM performance. We also consider several ground stations with varying properties (proximity to land, water bodies, and Earth's polar regions) to identify how geospatial properties affect the model's performance. Our results show ERSM works with a suitable resolution for MagNav with ground stations located several thousand kilometers from the desired prediction's location. 

%% file: src/introduction.tex
Navigation and localization are critical pillars for successfully operating human-controlled and autonomous vehicles. Most vehicles use a Global Navigation Satellite System (GNSS) for navigation. However, GNSS systems (such as GPS) are unreliable in adversarial settings where GPS jamming or spoofing can prevent successful localization \cite{altaweel2023gps}. Complementary navigational methods can be used to rectify this. Recently, magnetic anomaly navigation (MagNav) - using Earth's geomagnetic crustal anomaly field - has shown promise as a robust complementary navigation method leveraging existing environmental signals \cite{Canciani2016Positioning, Canciani2017}. 

Temporally varying components of the geomagnetic field (\textit{i.e.}, Earth's disturbance field), representing less than 5\% of the magnetic field, need to be filtered from scalar measurements to isolate the magnetic anomaly. The time-varying field can be estimated via an additional state in an extended Kalman filter (EKF) for fast-moving vehicles, such that temporal disturbances separate from spatial frequencies \cite{nielsen2023extended}. 

For slow-moving vehicles and vehicles revisiting the same area in a short period (\textit{e.g.}, for magnetic anomaly map making), the EKF approach is not applicable. In this case, practitioners require a ground reference station within 100 km of the vehicle. However, this is impractical for many situations, including marine track surveys where fixed ground stations are not possible. We propose the Extended Reference Station Model (ERSM) to redress this. ERSM applies a correction to an extended reference ground station (ERS) to predict the temporally varying scalar field at a local reference station (LRS) near a vehicle, marine track survey, etc. Broadly, ERSM is a step forward in enhancing the viability of MagNav to provide complimentary navigational methods in contested environments.  

\textbf{Key Contributions:} In this work, we consider the ERSM framework. \cite{Bergeron2023} considered a preliminary version of ERSM  in a pipeline for anomaly mapping via aeromagnetic surveys with a limited evaluation of ERSM. Here, we extend the framework to multiple regression models and evaluate it at several ground stations in various locations worldwide. 

Our evaluations consider how displacement (latitudinal and longitudinal) between the ERS and LRS affects model performance. We also evaluated how the models perform with and without a water body between them (to simulate a use case for marine track surveys). Across many of our results, we find a typical performance of less than 10 nT root mean square error (RMSE), with a median performance below 5 nT in several cases. While practitioners prefer error less than 1nT, ERSM's median performance is within an acceptable range of 10nT for anomaly map making.

%% file: src/prelim.tex
\textbf{Geomagnetic Field \& Temporal Variation: } The measurable geomagnetic field consists of several components, including the core field (1) and crustal (anomaly) field (2). We classify the remaining dynamic components as the temporal field (3).

\begin{enumerate}[leftmargin=*]
    \item \textbf{Core Magnetic Field ($\bm{\vec{B}}_\textbf{core}$):} The rotation of conductive fluids in Earth's outer core generates an electric field, driven by heat from the inner core, which produces the core magnetic field \cite{Hulot2011TerrestrialMagnetism}. Practitioners use the International Geomagnetic Reference Field (IGRF) core field model to estimate the core magnetic field component \cite{Alken2021}, and it is revised by the International Association of Geomagnetism and Aeronomy (IAGA) every five years to account for changes in the core field \cite{Hulot2011TerrestrialMagnetism, campbell2003introduction}. 

    \item \textbf{Crustal Magnetic Field ($\bm{\vec{B}}_\textbf{anomaly}$):} Permanent and induced magnetic fields generated by rock and mineral deposits cause the crustal magnetic field. Minerals below their respective Curie temperatures, where materials exhibit ferromagnetism \cite{Sabaka2002ComprehensiveModel}, cause the dominant permanent and induced fields. Earth's core field induces magnetic fields in minerals and rocks and creates the crustal (or anomaly) field. The crustal field represents a minor component ($1\% - 5\%$) of the measurable field and remains static, \cite{Canciani2016Positioning}.

    \item \textbf{Temporal Magnetic Field ($\bm{\vec{B}}_\textbf{temporal}$):} The temporal field includes magnetic fields caused by the ionosphere, the magnetosphere, and coupling currents \cite{Canciani2016Positioning, campbell2003introduction}. 
    The electrically conductive ionosphere produces Solar-Quiet (Sq) variations due to Earth's rotation. As Earth rotates, regions facing the sun heat up while regions facing away from the sun cool down, resulting in a diurnal (day-periodic) solar heating and cooling cycle which creates a time-varying geoelectric field. The geoelectric field, in turn, induces a diurnal magnetic field. Other temporally varying components within the ionosphere include the auroral current streams and equatorial electrojet (EEJ) streams, which depend on the season and time-of-day-dependent impact and are observed near the polar and equatorial regions, respectively \cite{Beck1969MagneticFields}.  

    Magnetic fields from Earth's magnetosphere result from charged particles emitted by the sun interacting with Earth's magnetic field. The interaction between the magnetosphere and the ionosphere forms coupled currents that induce a source of magnetic fields \cite{Kivelson1995}. This interaction is stronger near the poles, resulting in the Field Aligned Current (FAC) \cite{Sabaka2002ComprehensiveModel, Kivelson1995}, which further distorts Sq variations near Earth's poles \cite{campbell2003introduction}. 

    As the dominant component within ${\vec{B}}_\text{temporal}$ is the Sq field, we interchangeably refer to ${\vec{B}}_\text{temporal}$ as the diurnal variations (DV) due to its 1-day periodic behavior. 
\end{enumerate}

These components added together allow us to model our total field, 
\begin{equation}\label{eq:total_b_field}
    \vec{B}_\text{total} = \vec{B}_\text{core} + \vec{B}_\text{temporal} + \vec{B}_\text{anomaly}.
\end{equation}

\textbf{Magnetic Anomaly Navigation (MagNav): }The static nature of Earth's geomagnetic crustal field makes it a suitable signal for navigation. Furthermore, the magnetic signature is more resistant to signal jamming as magnetic dipole sources decay at a rate of $1/r^3$, where $r$ is the source-sensor separatin distance (as opposed to $1/r^2$ for RF jamming). Thus, navigation via crustal magnetic sources (MagNav) has shown promise, especially in GPS-contested environments \cite{Canciani2017}, with recent flights demonstrating its viability \cite{sansano2023mission, canciani2021magnetic}.

Successful navigation using magnetic anomaly maps requires isolation and removal of $\vec{B}_\text{core}$, $\vec{B}_\text{temporal}$, and the magnetic field produced by the vehicle \cite{Canciani2017} to retrieve the magnetic anomaly. While isolating $\vec{B}_\text{temporal}$ in high-speed vehicles is feasible with filtering techniques, it is challenging for low-speed vehicles \cite{Canciani2016Positioning}. Similarly, creating accurate magnetic anomaly maps with sufficient resolution requires removing the temporal component \cite{Bergeron2023}. 
To measure and remove the temporal field reliably, anomaly map-making and navigational algorithms require a fixed base station (within 100 km) of the vehicle's trajectory. 
However, it is infeasible in many situations to have a fixed base station nearby. To address this limitation, this work focuses on developing a model that extends the 100 km limit to predict the temporal magnetic field variations.

Currently, magnetometers sensitive enough to measure the magnetic anomalies are scalar magnetometers. Thus, in standard practice, magnetic anomaly maps and models are made using scalar magnetometers. Therefore, it is infeasible to estimate $\vec{B}_\text{temporal}$ by directly using Equation \ref{eq:total_b_field}.
To compensate, we make the following (practical) assumptions $|\vec{B}_\text{core}| \gg |\vec{B}_\text{anomaly}|$, $|\vec{B}_\text{core}| \gg |\vec{B}_\text{temporal}|$, and construct the following approximation: 
\begin{equation}
    |\vec{B}_\text{total}| \approx |\vec{B}_\text{core}| + |\vec{B}_\text{anomaly}| + |\vec{B}_\text{temporal}|.
\end{equation}
A detailed mathematical exposition on this approximation can be found in \cite{Bergeron2023thesis}. 
Using the above approximations and the following $|\vec{B}_\text{temporal}| > |\vec{B}_\text{anomaly}|$, we can remove the anomaly term, and state: 
\begin{align}
    |\vec{B}_\text{total}| \approx |\vec{B}_\text{core}| + |\vec{B}_\text{temporal}|,\label{eq:overall_approx}\\
    \intertext{or state, }
    |\vec{B}_\text{total}| - ( |\vec{B}_\text{core}| + |\vec{B}_\text{temporal}|)  =  B_\epsilon,
\end{align}
where $B_\epsilon$ is an unknown, static, scalar quantity; thus, in our model, we use Equation \ref{eq:overall_approx}. Magnetic anomaly mapping pipelines can estimate $|\vec{B}_\text{anomaly}|$ despite $B_\epsilon$ since the leveling process assumes and corrects for biases in the provided measurements. Similarly, proposed MagNav, state filters assume an inherent bias exists and remove it \cite{Canciani2016Positioning}. Hence, it is paramount to accurately predict the swing in temporal variations (as the 0 Hz component can be estimated).

\textbf{Prior Work: } Several existing models make predictions of the time-varying component of the Earth's magnetic field. The Dedicated Ionospheric Field Inversion\footnote{The latest version of DIFI is DIFI-8 and is available at \texttt{https://github.com/CIRES-Geomagnetism/DIFI}.} (DIFI) algorithm provides diurnal variations predictions using base station and satellite data for training \cite{Chulliat2013, Chulliat2016}. However, it has a low resolution of high-frequency components and does not fully capture the magnitude of the diurnal swings, and is not recommended for use beyond 2024, limiting its real-time deployment. This results in a high RMSE $> $ 120 nT without corrections and RMSE $> 14 $ nT with corrections, as shown in sample DIFI predictions (Figures \ref{fig:difi_sample_pred} and \ref{fig:difi_sample_pred_mean_adj}). The High Definition Geomagnetic Model (HDGM)\cite{hdgm2025, Nair_2021} makes predictions that include the main field and secular variation. HDGM can also include real-time components (HDGM-RT), which provide predictions that include both diurnal variations and the real-time disturbance field. Similar to DIFI, this suffers from the low resolution of higher frequencies. The performance of DIFI and HDGM also suffers during solar storms ( $K_p \geq4$) due to the underlying models not accounting for this information. To address these issues, the extended reference station model extends the viability of a base station beyond 100 km for magnetic anomaly map making and navigation.

%% file: src/ersm_method.tex
\label{ref:methods}
\subsection{Methodology Overview}

\begin{figure}
    \centering
    \includegraphics[width=\linewidth]{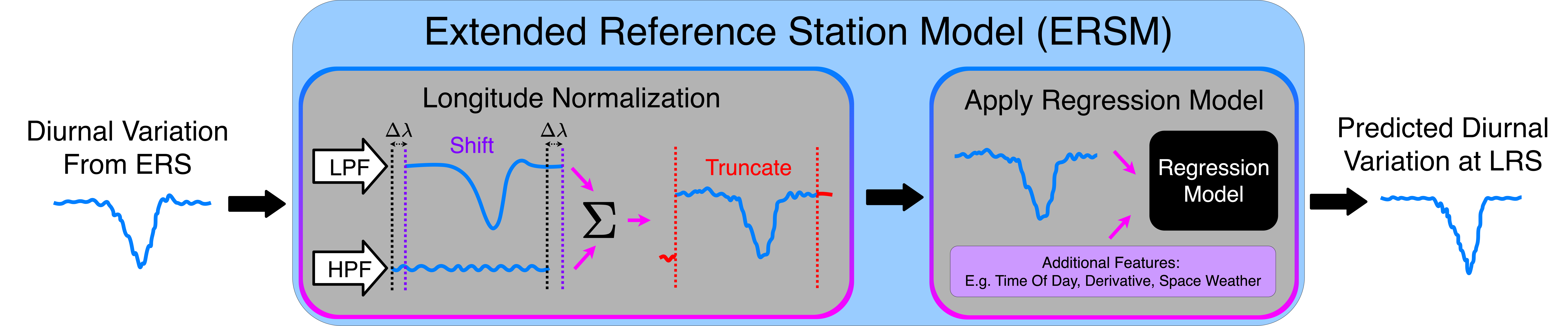}
    \caption{The Extended Reference Station Model (ERSM) inference pipeline. At a high level, ERSM has two components. 1) Longitude normalization is achieved by using a high-pass filter (HPF) and a low-pass filter (LPF) to shift low-frequency components to account for the Earth's rotation. 2) A regression model to account for shape and magnitude variations due to local geographic features. We also truncate the LRS's DV to ensure timestamp alignment during training of the regression model.}
    \label{fig:model_pipeline}
\end{figure}

The Extended Reference Station Model (ERSM) predicts the local geomagnetic diurnal variation (DV) at a local reference station (LRS) by applying a correction derived from measurements at an extended reference station (ERS) located up to several thousand kilometers away. The correction is an \textit{a priori} learned relationship between the ERS and LRS in ERSM. It is established by computing and applying a low-frequency offset to the extended reference station to account for longitudinal differences between locations and then applying a regression model to account for location-dependent differences between the two reference stations, as seen in Figure~\ref{fig:model_pipeline}. We evaluate in Section \ref{sec:results} how the choice of ERS affects the performance of the ERSM, considering how temporal variations in the magnetic field depend upon latitude\cite{campbell2003introduction}.

We implemented and evaluated linear, $k$-nearest neighbors (kNN), and neural-network regression models within the regression-agnostic ERSM framework. The linear regression model takes the form $\hat{y} = ax + b$ where $a$ and $b$ correspond to the scale and offset parameters, respectively. The linear model formed a baseline for additional regression models\cite{Bergeron2023}\cite{Mammal2025}. The kNN regression is a non-parametric regression model that approximates non-linear and unknown relationships by considering the $k$ closest neighbors from training data, as evaluated by a distance metric.

We included the kNN model because it uses prior data to make a prediction, thus making it suitable for the quasi-periodic diurnal variation. Lastly, for the neural network, we use an ensemble of neural networks with an architecture inspired by ResNets to create our model \cite{he2016deep}. We considered this model as it overcame a drawback of the kNN model (its restriction to the range of points provided as training data) and made predictions outside of the training data range.

\subsection{Longitudinal Normalization}
Longitudinal normalization accounts for phase shifts of the low-frequency component of the diurnal variations at the extended and local reference stations due to Earth's rotation \cite{Bergeron2023}. The frequency correction is based on the longitudinal difference between the ERS and LRS since the rotation is primarily longitudinal. To apply longitudinal normalization, we use
\begin{equation} \label{eq:delta_long}
\Delta\lambda = \lambda_{e}-\lambda_{l}
\end{equation}
where $\lambda_e$ and $\lambda_l$ correspond to the longitude of the extended and local reference stations, respectively, to compute the difference in longitude. Then, using
\begin{equation} \label{eq:time_offset}
t_{l} = \frac{\Delta\lambda}{\omega_{\text{earth}}}
\end{equation}
where $\omega_{\text{earth}} = 0.004178 \deg/\text{sec}$ is the angular frequency of the Earth's rotation, and $t_{l}$ is the time offset in seconds, we compute the time offset.

After computing the time offset, we separate the low and high frequencies of the ERS's DV using low and high-pass Butterworth filters, \cite{Butterworth1930}, with cutoff frequencies of $0.33~\text{cycles/hour}$. 
We time-shift the low-frequency component by the time offset and add it to the high-pass component.
The result is the time-aligned extended reference station. The time shift truncates the time of the longitude-normalized extended reference station data. We also truncate the data collected in the LRS to prevent timestamp misalignment between the LRS and ERS. Once the extended reference station is time-aligned, we train a regression model to learn a relationship between the time-aligned stations. We describe the regression approaches we considered below.

\subsection{Linear Regression}
The linear regression approach minimizes the residual sum of squares (RSS), 
\begin{equation} \label{eq:RSS}
\text{RSS}(a,b) = \sum_{t}(B_l[t]- (a\x B_{eta}[t] +b))^2
\end{equation}
where $B_l[t]$ and $B_{eta}[t]$ correspond to the temporal variations at the LRS and time-aligned ERS to compute the optimal (least squares) scale and offset: $a_{ls}$ and $b_{ls}$. Then, at inference time, $a_{ls}$ and $b_{ls}$ are used to predict the disturbance field by the LRS from the ERS data. 

\subsection{k-Nearest Neighbors Regression}
\begin{figure}
    \centering
    \includegraphics[width=1.0\linewidth]{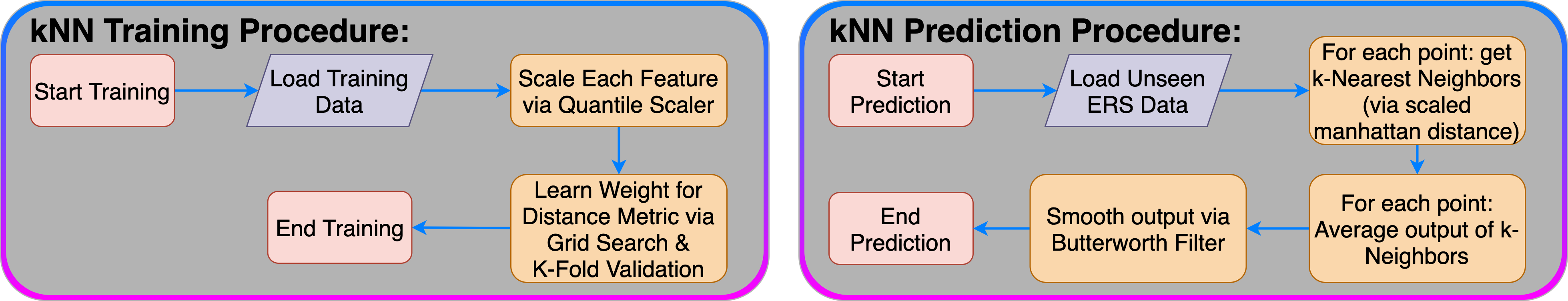}
    \caption{The procedure for training and predicting with the kNN regression in ERSM. During training, we scale the features, tune the hyperparameter for the distance metric, and use this distance metric to make predictions.}
    \label{fig:kNNFlow}
\end{figure}
\textbf{Feature Selection and Preprocessing:} The kNN model uses three features: the time of day, the time derivative of the time-aligned extended reference station data, and the time-aligned ERS data. We compute the time of day (TOD) using 
\begin{equation} \label{eq:tod}
\hat{t} = t\,\,\mathrm{mod}(24\x3600)
\end{equation}
where $\hat{t}$ corresponds to a time of day in seconds, and $t$ is the input the time in seconds. The standard used for time in seconds can be selected by the user, and in our implementation, we use Unix time. We included the TOD feature because it allows the model to make time-aware predictions by assuming the diurnal variation is quasi-periodic. To compute the time derivative of the time-aligned ERS data we used a finite time difference approximation (\textit{i.e.}, $\frac{B_{eta}[t+1]-B_{eta}[t]}{\Delta t}$, where ${\Delta t}$ is the timestep of the timeseries), and we found it improved predicting the peaks and troughs of the diurnal variation. 

Before adding the data to the kNN regression, a quantile transformation converted the data for each feature to a uniform distribution \cite{pedregosa2011scikit}, and a scale factor was applied to each feature to ensure that each feature had a similar mean and variance of points. In practice, this approach worked better than applying a minimum-maximum scaler that linearly scales data into a specific interval (\textit{e.g.} [0,1]) since it does not get skewed by outliers. 

\textbf{Distance Metric Selection: }After experimentation, we found that the weighted Manhattan distance metric
\begin{equation} \label{eq:dist}
d(x,x') = |x_{t} - x'_{t}| + \alpha|x_{d} - x'_{d}| + |x_{m} - x'_{m}|
\end{equation}
where $x,x'\in \mathbb{R} ^3$ contain $x_t$, $x_d$, and $x_m$, which correspond to the TOD, time derivative, and scalar ERS  geomagnetic DV, respectively. $\alpha > 1$, the optimal weight to apply to the time derivative feature, is constrained to be greater than one to place greater importance on the TOD and magnitude features and prevent the model from overfitting to the time derivative feature.

\textbf{Hyperparameter Tuning:} We used a grid search method to determine the optimal $\alpha$ value and $k$, the number of neighbors to consider, based on the root mean square error (RMSE). We constrain $1 \leq \alpha \leq 50$ for the grid search to ensure that the time-derivative feature remains relevant. To further fine-tune $\alpha$ on a continuum, we used basin hopping optimization, since the grid search only considers integer values of $\alpha$ \cite{bassinHopping}.

To cross-validate $k$ and $\alpha$ during the grid search, we used K-fold cross-validation, where each fold corresponded to one day's worth of data. In K-fold validation, we hold out one fold to assess the model's RMSE while training the model on the remaining folds. This process repeats for each fold, allowing us to report the average performance across all folds. To prevent magnetically noisy time frames from skewing results in the cross-validation, we used the $K_p$ space weather index to prune times with $K_p > 4$ from the test fold (\textit{i.e.}, the data held as validation). Since this causes each fold to have a varying number of data points, we employed a weighted average across the folds to determine the model's performance.  

\textbf{Prediction:} After hyperparameter tuning, the LRS's DV can be predicted using the kNN regression. The kNN model makes the prediction using
\begin{equation} \label{eq:pred}
P(x) = \sum_{y\in N_{k}(x)}w(x,y)\x y_\text{out}
\end{equation}
where $w(x,y)$ is the weight function and $y_\text{out}$ is the known output corresponding to the points $y \in N_k(x)$, the k-closest neighbors to the input. During inferencing, we used an inverse-distance weight function
\begin{equation} \label{eq:weight}
w(x,y) = \frac{\frac{1}{d(x,y)}}{\sum_{z\in N_{k}(x)}\frac{1}{d(x,z)}}.
\end{equation}
To denoise the model's output, we applied a low-pass Butterworth filter with a 1.5 cycles/hour cutoff to the output. A summary of the kNN model can be seen in the flow chart in Figure \ref{fig:kNNFlow}.

\subsection{Neural Network Model}
\begin{figure}
    \centering
    \includegraphics[width=0.7\linewidth]{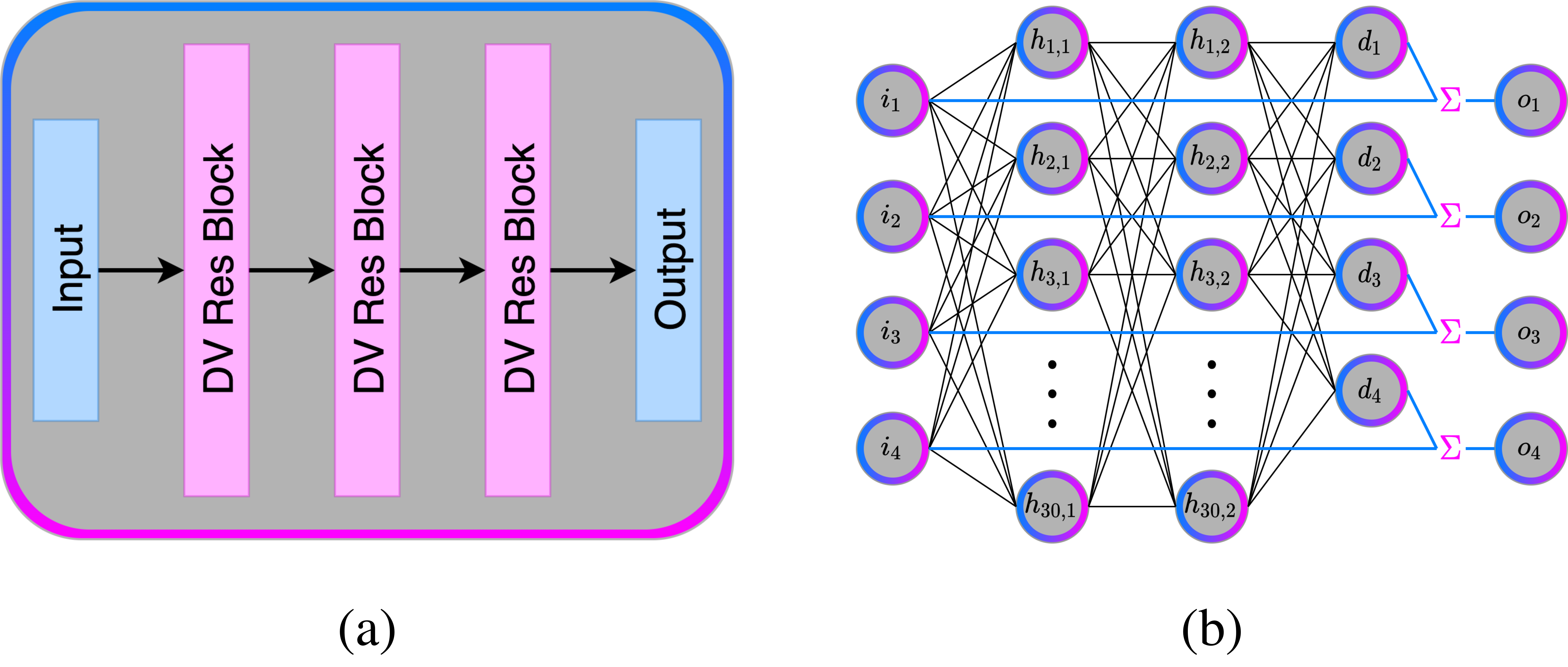}
    \caption{\label{fig:res_block} The neural-net model we implemented in ERSM. (a) The high-level representation of the architecture using 3 ERSM residual blocks. (b) Lower-level representation of each ERSM residual block. The block's output is the sum of the input and the output of the last linear layer.}
\end{figure}

\textbf{Network Architecture: }The neural network architecture, Figure \ref{fig:res_block}(a), adopts skip connections to add a block's inputs to the block's output, which consists of two fully connected layers, as first done in the ResNet model \cite{he2016deep}. Figure \ref{fig:res_block}(b) shows the architecture for a given block. The advantage of this structure is that it preserves the linear relationship between the locations during the day by gradually adding non-linearities without entirely distorting the initial information passed to the model.


    

\begin{figure}
\centering
\includegraphics[width=.4\linewidth]{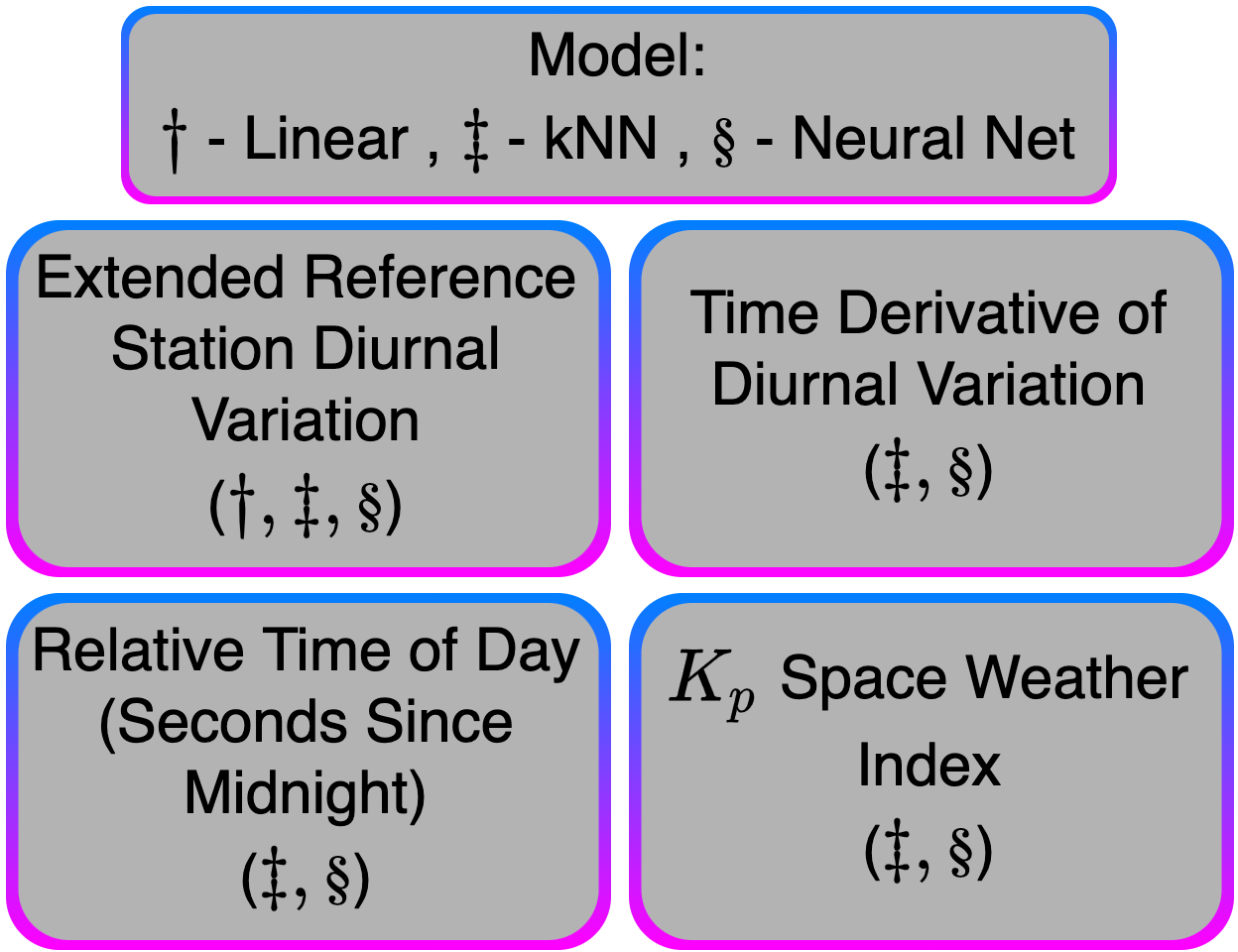}
\caption{Features used in each model.}
\label{fig:feature_selection}

\end{figure}

Our architecture has 3 such blocks with 30 nodes per hidden layer in each block. Then, we used a linear layer to take the 4 outputs from the last block to make a prediction. We used an ensemble of 16 such models during inference to enhance the robustness of this model. We applied a low-pass filter to denoise the model, similar to the kNN model, and subsequently averaged the output of each model to make a prediction. During averaging, we removed the output corresponding to the point farthest from the median prediction, which was assumed to be an outlier. Averaging occurs over 15 ($=16-1$) points. The averaging strategy makes the model robust to outliers created by models diverging from the optimal solution.
We also used batch normalization and dropout layers during training to prevent the model from overfitting \cite{ba2016layer, srivastava2014dropout}.

\textbf{Feature Selection: }Alongside the three features directly used for a KNN, the $ K_p$ index is a fourth feature used for the neural net model. 
We found that using the $K_p$ index to prune the training data for the neural net model provided good results. Instead of using the $K_p$ index (like the kNN model), we found that using a boolean value to indicate that a $K_p$ was greater than or equal to 7 produced better results. The underlying intuition of this approach is that it allows the model to distinguish noisy time frames from time frames with limited noise without overfitting to the $K_p$ index. A summary of the feature selection for all the models can be seen in Figure \ref{fig:feature_selection}.

%% file: src/exp_method.tex
\textbf{Implementation:} To implement the models, we used the \texttt{MAMMAL}\footnote{The github for \texttt{MAMMAL} is \texttt{https://github.com/PowerBroker2/MAMMAL}.} library, \cite{Bergeron2023thesis}, which provides data processing tools for INTERMAGNET data. The baseline linear regression is the only model available in \texttt{MAMMAL}.
We used Python's Sklearn to implement the KNN regression and the quantile transform (and the required data pre-processing) \cite{pedregosa2011scikit} and \texttt{PyTorch} to implement the neural network models \cite{paszke2019pytorch}.

\textbf{Datasets: } To curate the datasets, we gathered geomagnetic data from INTERMAGNET \cite{Intermagnet2021-fd} and space weather data collected by GFZ \cite{Matzka2021-zn}. INTERMAGNET provides users with data sampled at different rates (once per minute and once per second) and of varying quality. Our study uses data with the 'best available' quality sampled once a minute. Minute-sampled data provided sufficient granularity and kept the model training time at a minimum (particularly for the kNN and Neural Network models). After obtaining the INTERMAGNET data, we pre-processed it by removing outliers and linearly interpolating missing points using the \texttt{MAMMAL} library. Then, the core field was subtracted from it using the IGRF core field model \cite{Alken2021} to isolate $|\vec{B}_\text{temporal}|$ according to Equation \ref{eq:overall_approx}. 

The space weather ($K_p$) data was sourced directly from GFZ's webpage, along with its timestamps. The $K_p$ value was averaged across 13 stations to estimate the space weather globally \cite{Matzka2021-zn}. The $K_p$ data provided is updated hourly. To adjoin the $K_p$ value to the minute-sampled INTERMAGNET data, we used the most recent $K_p$ value available (as opposed to interpolation schemes that used future data).

\begin{figure}
    \centering
    \includegraphics[width=0.9\textwidth]{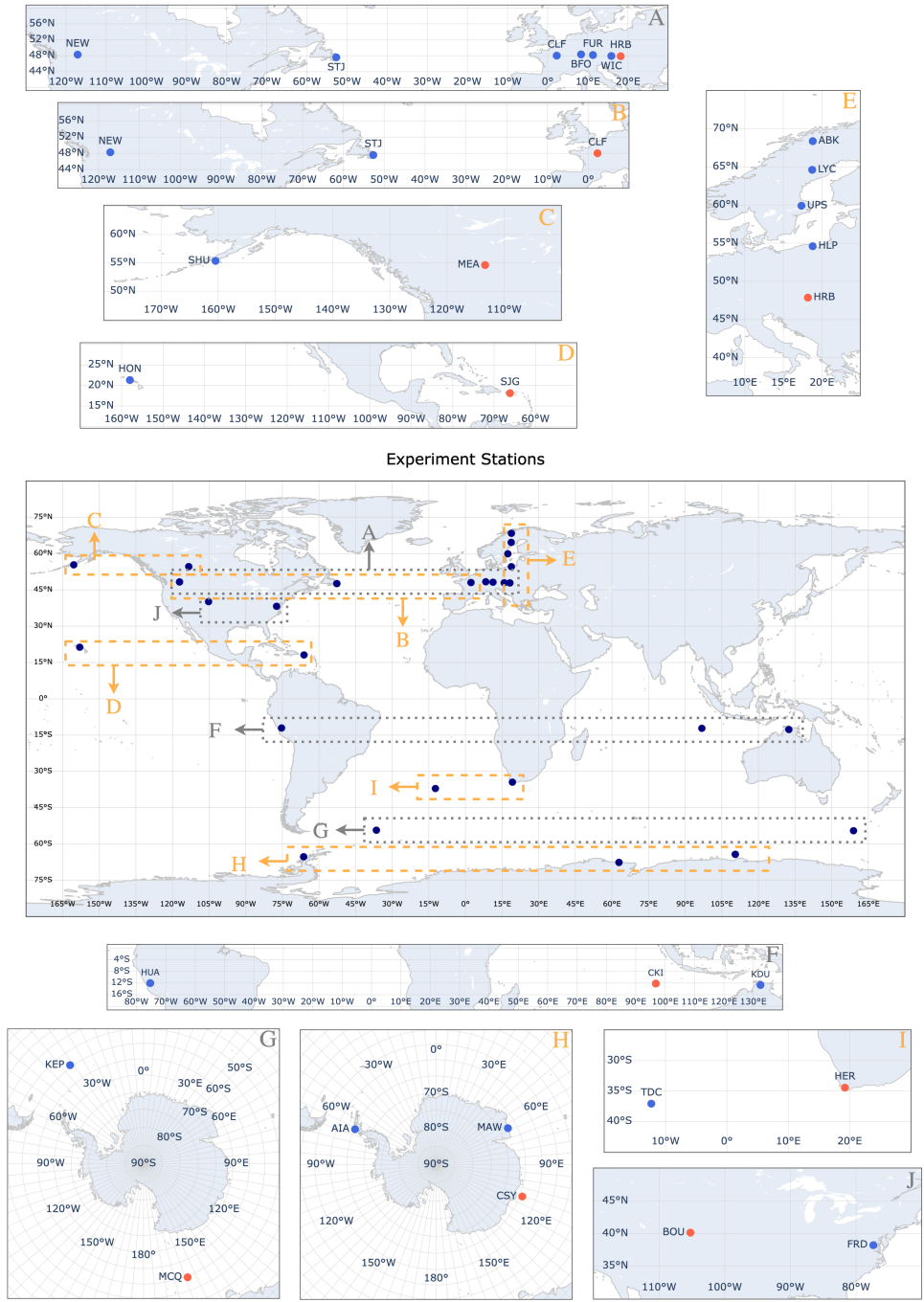}
    \caption{The stations we used to evaluate ERSM. The central figure (Experiment Stations) provides a global view of the locations of the stations (dark blue dots) and how we selected a diverse range of stations. To identify the experiment in the central figure, we provide an alphabetical identification for each group of stations used in an experiment. The remaining figures show each experiment (with the alphabetical coding). In these figures, the red dot and light blue dots correspond to the LRS and ERSs, respectively. For experiments G and H, we provide the orthographic projections to illustrate the proximity between stations more clearly.}
    \label{fig:experiment_locations}
\end{figure}

\textbf{Experiment Selection: }To test the models, we used data from several INTERMAGNET ground magnetometer stations, shown in Figure \ref{fig:experiment_locations}. The stations were clustered based on having similar latitudes, longitudes, or shared properties (i.e., proximity to a water body or polar regions) to categorize the performance of ERSM based on the ERS and LRS selection. To explore the effect of station separation, we selected stations to ensure that our analysis included results over various station separation distances (e.g., 173 km to 17,173 km). 

\textbf{Training \& Evaluation: }
To train the three models (linear, kNN, and neural-net), we used 10 days of data and followed the respective training paradigms detailed in Section \ref{ref:methods}. The model used the following 7 days to evaluate the prediction made by ERSM. We used 22 such 17-day blocks between August 1$^{\text{st}}$, 2022 and August 9$^{\text{th}}$, 2023 ($22*17 = 374$ days) for each experiment. 

We used the Root Mean Square Error (RMSE) metric to quantify the model's performance. A higher RMSE indicates worse performance, while a lower RMSE indicates better performance. In our results, we present the daily RMSE marginalized over the 22 training-evaluation blocks. Using these results, we can identify ERSM's general performance instead of its seasonal performance.\footnote{A handful of stations were missing data on certain days. If a missing day corresponded to training data, we trained with fewer data points. If the missing day was an evaluation day, we reduced the evaluation time from 7 to 6 days to maintain the training blocks. Occasionally, data was missing for several weeks. In this case, the number of blocks used decreased.}  

%% file: src/results.tex


\begin{figure}
    \centering
    \includegraphics[height = 1.75in]{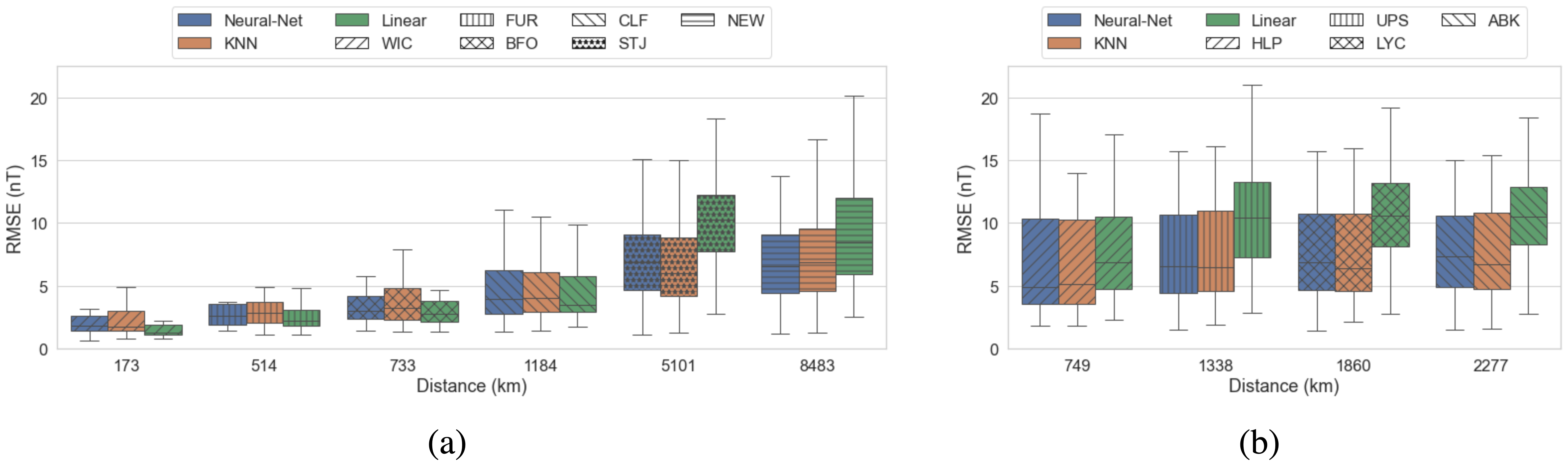}
    \caption{\label{fig:geog_sep}ERSM performance tested against geographic separation between the ERS and LRS. For both longitudinal (a) and latitudinal (b), we fixed the LRS as HRB, and varied the distance from the ERS to the LRS. In both cases, we observe worse performance at farther distances. The hashing pattern in the figure can identify the station name at each distance.}
\end{figure}

\subsection{Geographic Separation on Model Performance} \label{sec:results:sep}

To quantify the effect of station separation, the models were evaluated using stations with longitudinal separation (and minimal latitude separation). The 10 days of training data from the stations in Figure \ref{fig:experiment_locations}(A) were used to predict the following 7 days where the local reference station (LRS) was Hurbanovo, Slovakia (HRB). The results in Figure \ref{fig:geog_sep}(a) demonstrate a decay in performance as the longitudinal separation increases. For the linear model, there is a sharper decay after the CLF ground station. However, the neural network and kNN models perform better with a median RMSE around 7.5nT when using STJ as the ERS (a distance of 5101 km). The decay in performance (particularly for the linear regression) is attributed to non-linear components of the relationship between Diurnal Variations being more profound at greater distances. 

Conversely, at closer distances, the linear model performs better because the non-linearities are minimal, thus causing the kNN and Neural-Net approaches to overfit during training. This result supports the notion (and current practice), that limited to no correction would be needed (and used) for stations closer than 100 kilometers \cite{Bergeron2023}. 

We used the stations in \ref{fig:experiment_locations}(E) to gauge the effect of the latitudinal separation between stations.
For this experiment, HRB was again used as the local reference station. The results in Figure \ref{fig:geog_sep}(b) again show a decay in performance as station separation increases. Here, the neural-net and kNN approaches work better at greater latitudinal displacement due to there being more significant non-linearities in the relationship \cite{campbell2003introduction}.
Figure \ref{fig:geog_sep}(b) also demonstrates, an accelerated decay in performance when compared to the decay in performance in Figure \ref{fig:geog_sep}(a). This suggests that the models work better when the latitudinal separation is minimal.

\subsection{Station Performance across land mass}
\begin{figure}
    \centering
    \includegraphics[width=0.9\linewidth]{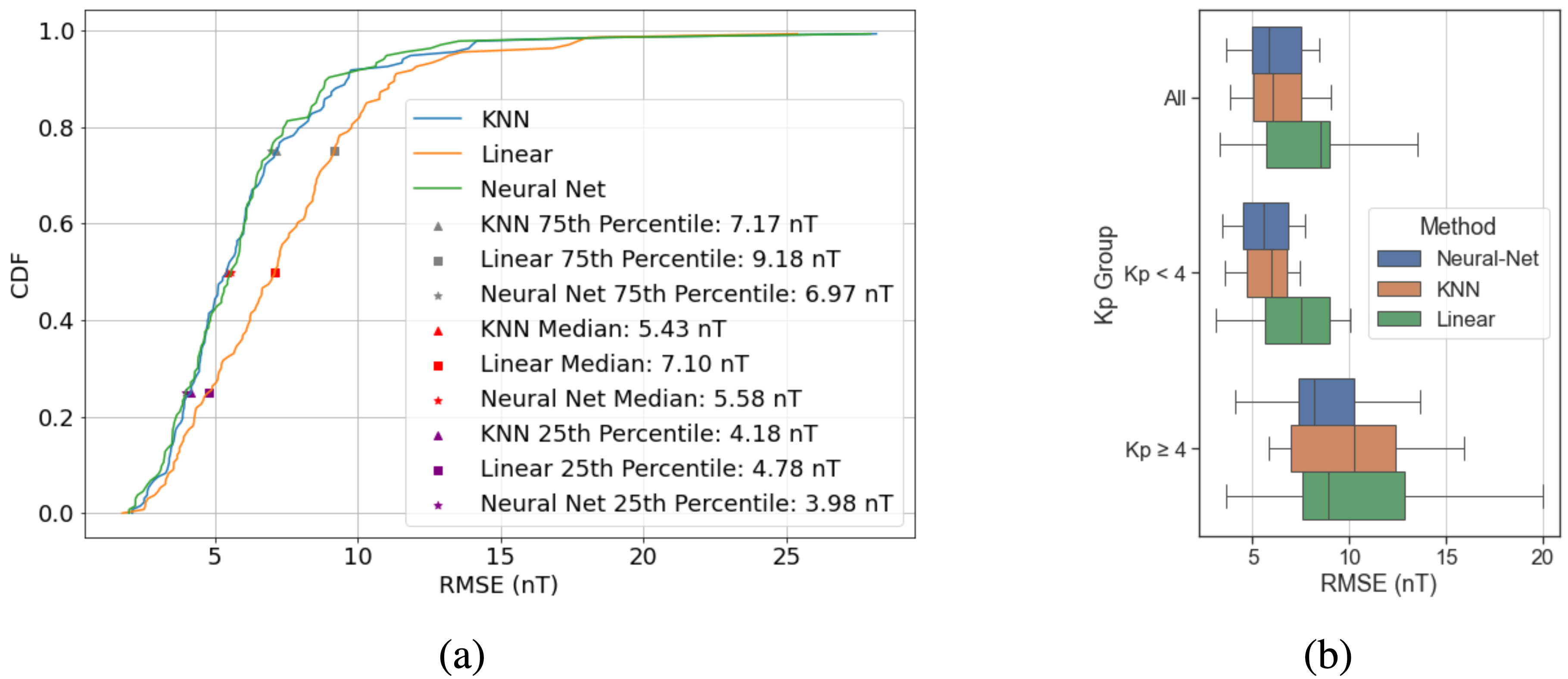}
    \caption{\label{fig:geog_bou_frd}ERSM performance between FRD and BOU. The kNN and Neural-Net models outperform the linear model (according to the CDF), with the Neural-Net model performing better under more severe space-weather conditions. (a) The CDF of model performance between FRD and BOU. (b) The space weather performance for each model from BOU to FRD.}
\end{figure}

We used stations in Fredericksburg, Virginia (LRS), and Boulder, Colorado (ERS) from Figure 5\ref{fig:experiment_locations}(J) to evaluate how the stations perform when separated by a landmass (with no ocean in between).
The stations are 2400 km apart and have a slight latitudinal separation (within $3\deg$). The CDF in Figure \ref{fig:geog_bou_frd}(a) distributionally illustrates how each model performs. The neural-net model and kNN model perform similarly, while the linear model's performance decays at the 50th percentile. 





\begin{figure*}
    \centering
    \includegraphics[width=\linewidth]{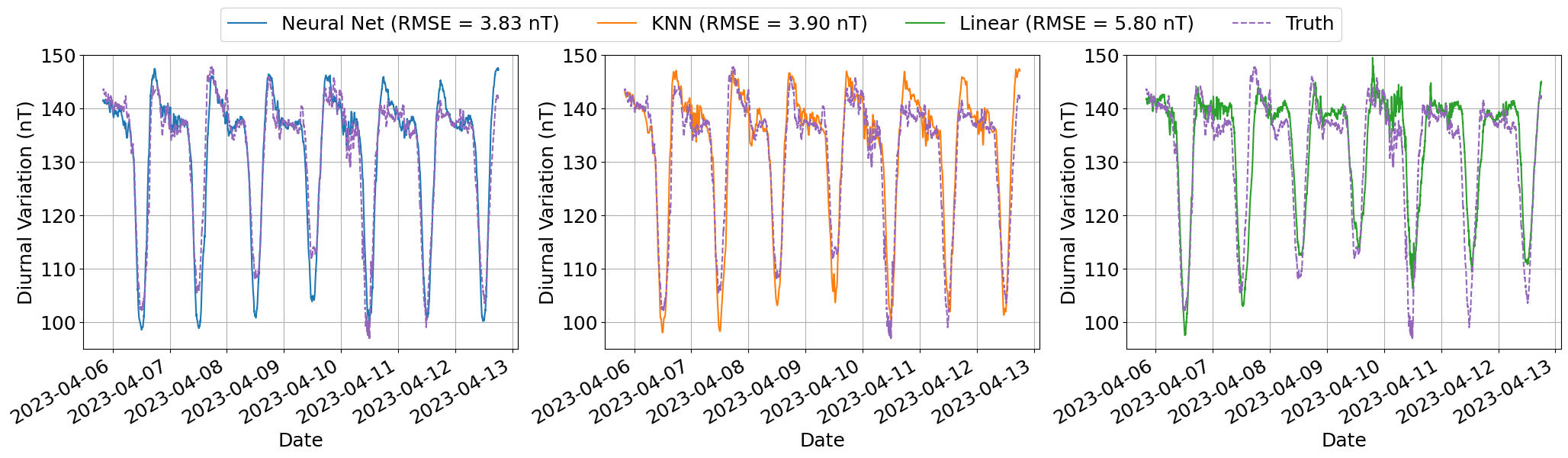}
    \caption{Sample prediction from 4/06/23 to 4/12/23 of each model. The kNN and Neural-Net models improve the alignment of the peaks and troughs of the DV with the ground truth to reduce the RMSE.}
    \label{fig:sample_pred}
\end{figure*}

To understand the performance gain achieved by the neural-net and kNN models, consider a sample prediction from the ERSM regression models in Figure \ref{fig:sample_pred}. As seen in the sample prediction, the kNN and neural-net models time align the troughs (using the time-of-day feature) as observed with troughs of the ground-truth matching the predictions. However, the troughs of the linear model have a time delta from the ground truth. The linear model relies solely on the longitudinal normalization for the time alignment. However, the kNN and neural-net models refine the longitudinal normalization to improve the performance. 

We also consider how each model performs under varying space weather conditions. Since $K_p \geq 4$ indicates a solar storm, we compared the model performance when $K_p \geq 4$ and  $K_p< 4$. Figure \ref{fig:geog_bou_frd}(b) illustrates that the neural-net model performs better than the kNN model under solar storm conditions. This behavior is attributed to the neural net learning a relationship, while the kNN model is unable to predict output values that it has not seen in the past. 


\subsection{Impact of a Water Body on Station Performance}

\begin{figure}
    \centering
    \includegraphics[width=6in]{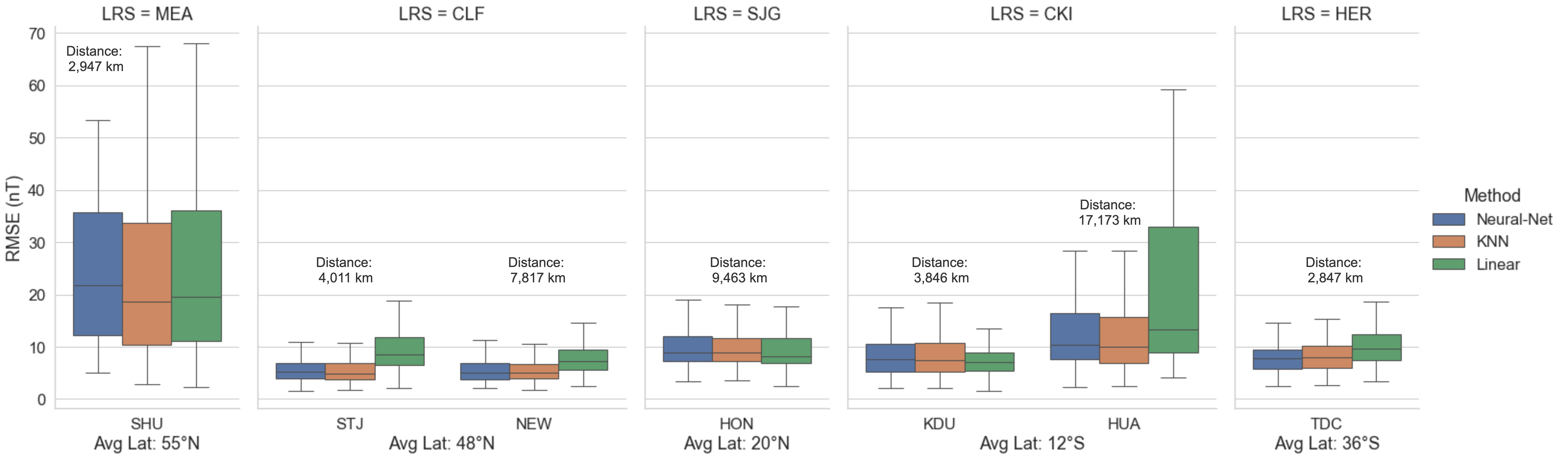}
    \caption{ERSM model performance across a water body. Data from 5 distinct latitudes (\textit{i.e.}, 5 distinct LRSs) is included. We find the median performance to be below 10nT for several models. We observe a decay in performance (higher RMSE) as station separation increases.}
    \label{fig:water_body}
\end{figure}

Marine-track surveys (and aeromagnetic surveys conducted over water bodies) require ground stations as well; however, it is impractical to have fixed (permanent) ground stations over water bodies. Thus, the ERSM framework can be applied in these surveys using a permanent ERS fixed to a land mass. In this regard, we ran experiments (similar to above) using stations where a water body was present between an ERS and an LRS. We stratified the stations by latitude to form several experiments (B, C, D, F, and I from Figure \ref{fig:experiment_locations}).


The results in Figure \ref{fig:water_body} show no general trend as to which model performs best. When considering the station separation (between the ERS and LRS), it becomes clear that further separation leads to worse performance. 

Notably, the models perform well for the experiments conducted around $48^\circ$N, despite the 4,011 km and 7,817 km station separation. 
However, the experiments run around 20$\deg$N, 12$\deg$S, and 36$\deg$S perform worse, suggesting the decay of model performance due to increased station separation varies based on latitude. Specifically for the experiments run around 48$\deg$N, we find a marginal decay in performance compared to the decay for stations around $12^\circ$S. The result by 12$\deg$S can be attributed to a greater intensity of the EEJ stream near the Indian Ocean \cite{onwumechili1998equatorial}. Additionally, the neural net and kNN models perform better throughout these experiments. However, there is no clear winner between the neural net model and the kNN model. Conversely, the linear model performs the worst in most water body experiments.

Lastly, the model performance using data around 55$\deg$N is noticeably worse than the remaining latitudes, thus motivating a study of ERSM near the Earth's geographic poles. In the following section, we assess the performance of ERSM using INTERMAGNET stations in Antarctica.

\subsection{Model Performance in Antarctica}
\begin{figure}
    \centering
    \includegraphics[width=0.5\linewidth]{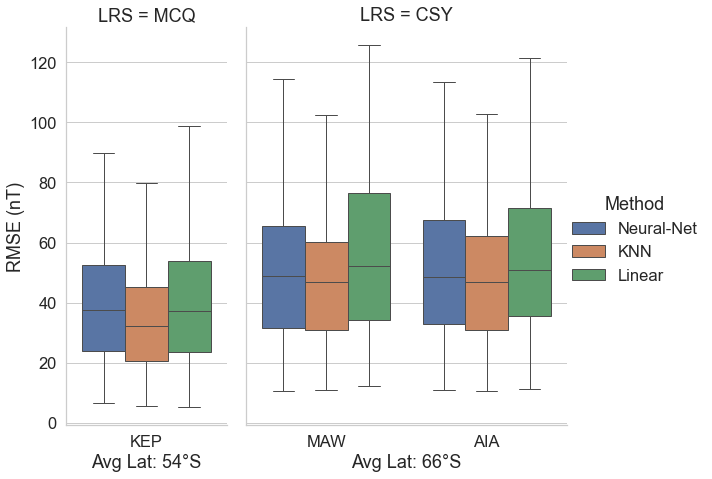}
    \caption{Model performance for each of the ERS over Antarctica.}
    \label{fig:antarctica_body}
\end{figure}

In Figure \ref{fig:antarctica_body}, the performance of each model is degraded (median error above 40 nT for most models) compared to prior experiments, where the median error is below 10 nT in most cases. The performance is comparable to the experiment from SHU to MEA (circa $55^\circ$N) in Figure \ref{fig:water_body}. Thus, we find the performance towards the geographic poles is worse. Additionally, the geomagnetic poles are close to these experiments; therefore, increasing proximity to these poles also degrades the results.

We attribute the poor model performance to magnetospheric processes affecting the geomagnetic temporal variations at higher latitudes near polar regions (i.e. $|\text{latitude}| > 60\deg$) \cite{campbell2003introduction}. The magnetospheric processes reduce the number of solar-quiet days observed, making it difficult to learn and predict the noisy diurnal variation. Similarly, the polar region electrojets induce high magnitude auroral currents, making it further challenging to predict the magnetic field within polar regions \cite{Beck1969MagneticFields, Canciani2016Positioning}. However, the effect of these processes is minimal in non-polar regions, which allows ERSM to perform better at mid-level and lower-level latitudes.


%% file: src/conclusion.tex
This work demonstrates that practitioners can apply the ERSM framework to ERSs to predict local diurnal variation and disturbances to the geomagnetic field. Our results illustrate that ERSM performs with suitable accuracy below 10 nT in most cases, with median performance below 5 nT in several experiments - particularly at mid-level and low-level latitudes. For practitioners, we recommend selecting extended reference stations with similar latitudes to the location of the desired prediction (as observed with our results in Figure \ref{fig:geog_sep}. We also find ERSM to be a viable option for marine track surveys, where having a fixed ground station is impractical. Thus, using ERSM, we reduce the dependence on a fixed ground station for vehicular surveys and navigation. 

\textbf{Limitations \& Future Work: }ERSM, as currently designed, cannot forecast beyond the timestamps of the input data from the ERS. Although this is not an issue for map-making, where ERSM can be applied post-hoc, for real-time navigation, it is. To account for this, we can develop forecasting strategies by forecasting the input ERS, forecasting the output from ERSM, or selecting an ERS east of the LRS. The latter option would use the longitudinal normalization step to shift the low, dominating frequencies to create future predictions while using a forecasting model to predict the less dominant, high frequencies. 

Lastly, ERSM can have high utility with portable magnetometer ground stations. Model training requires a temporary, portable ground magnetometer station. Afterward, a user can remove it and place it elsewhere. This approach is helpful for extensive aeromagnetic surveys covering several thousand kilometers, requiring several base stations, and for blue water marine-track surveys where fixed ground stations are unavailable. In this regard, future extensions of this work can include further field testing using portable stations to evaluate ERSM where fixed stations are unavailable. 

%% file: src/appendix.tex
The Dedicated Ionospheric Field Inversion (DIFI) chain model is used to predict Sq variations at mid- and low-latitude levels. The DIFI model uses data collected by the Swarm satellite mission (launched by the European Space Agency) to compute the coefficients of the basis functions that model the underlying geophysical properties of temporal variations within the ionosphere \cite{Chulliat2013, Chulliat2016}. Similarly, the Extended DIFI (xDIFI) model uses Swarm satellite and Challenging Minisatellite Payload (CHAMP) satellite data to model the Sq field. 

To compare the DIFI models against our ground truth, we take the vector Sq Field estimate provided by DIFI and add the vector core field (using the IGRF model). We then compute the magnitude of the core-adjusted DIFI prediction and subtract the magnitude of the core field to produce a DIFI sample prediction. Mathematically, this procedure can be represented as, 
\begin{equation}
    B_\text{DIFI}[t] = |\vec{B}_\text{DIFI}[t] + \vec{B}_\text{core}| - |\vec{B}_\text{core}|,
\end{equation}
where $\vec{B}_\text{DIFI}[t]$ is the original vector DIFI signal. We take this approach since our ground truth is computed by taking the scalar measurement from a ground station and subtracting the scalar core field. 

\begin{figure*}
    \centering
    \includegraphics[width=\linewidth]{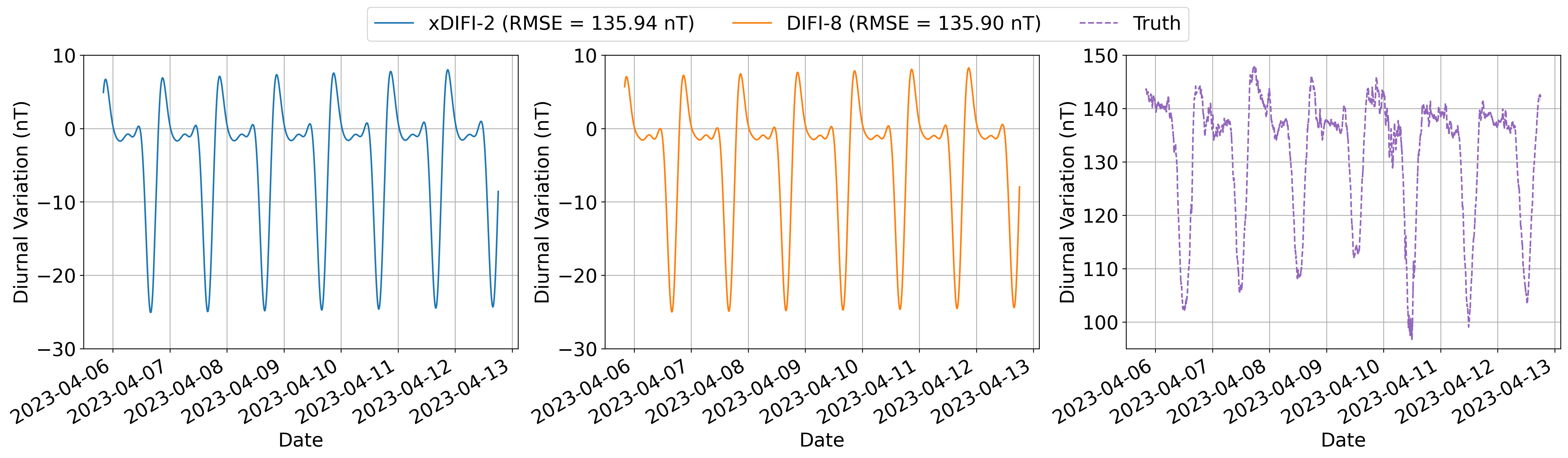}
    \caption{Sample prediction provide by DIFI-8 and xDIFI-2 from 4/06/23 to 4/12/23. The predicted DVs differ substantially from the ground truth, with an RMSE of more than 120 nT. }
    \label{fig:difi_sample_pred}
\end{figure*}

In Figure \ref{fig:difi_sample_pred}, we provide sample predictions from DIFI-8 and xDIFI-2. The raw RMSE compared to our ground truth is greater than 120 nT (making it unsuitable for directly predicting DV in magnetic anomaly mapping or navigation).

\begin{figure*}
    \centering
    \includegraphics[width=\linewidth]{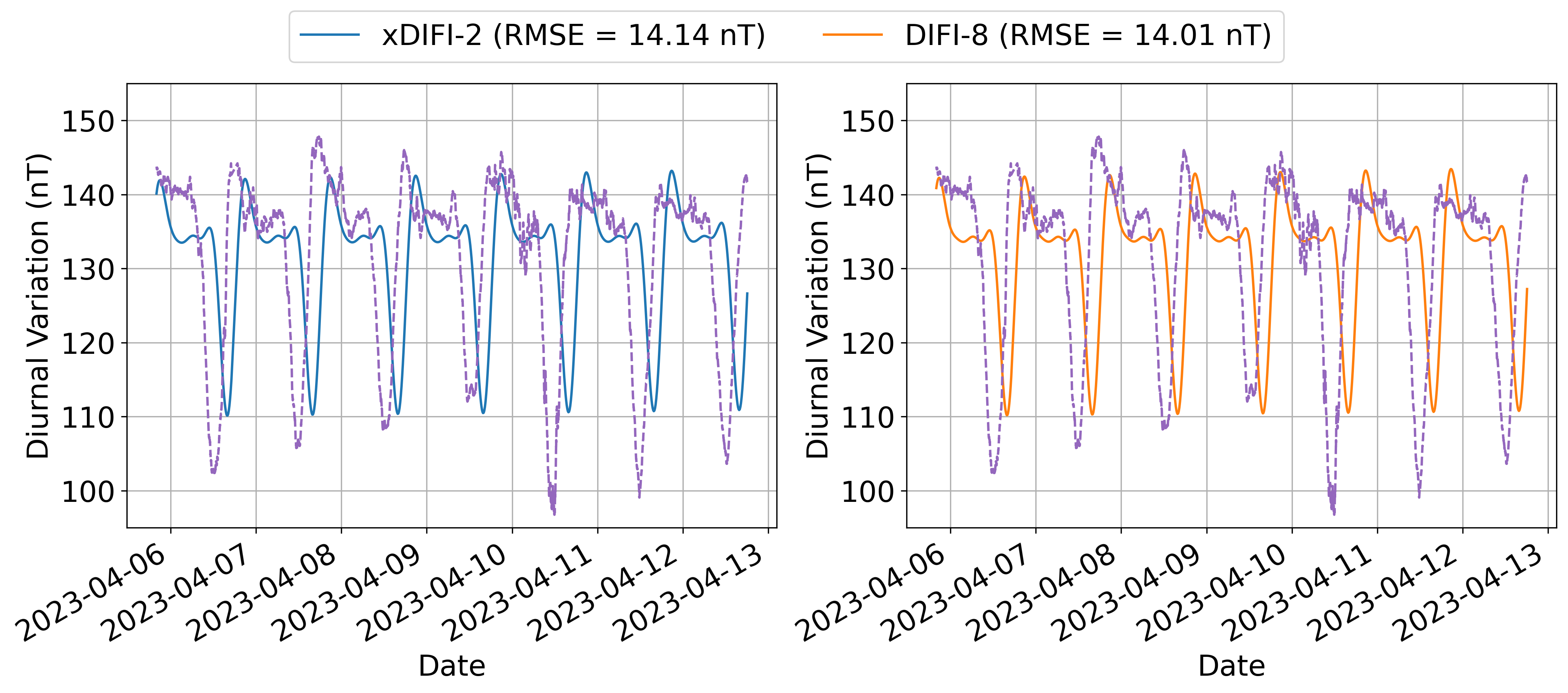}
    \caption{The sample predictions made by DIFI-8 and xDIFI-2 with the mean difference between the prediction and ground truth removed. We observe an improved RMSE compared to Figure \ref{fig:difi_sample_pred} - thus demonstrating the shape and size of the DIFI predictions have insufficient resolution for MagNav.}
\label{fig:difi_sample_pred_mean_adj}
\end{figure*}

As discussed in Section \ref{sec:prelim}, proposed state filters used in magnetic anomaly mapping and navigation can remove constant biases. Thus, to evaluate the suitability of DIFI, we offset the DIFI prediction by a fixed quantity, $b$, to minimize the RMSE between the adjusted DIFI prediction and the ground truth (\textit{i.e.}, $B_\text{adj}[t] = B_\text{DIFI}[t] -b$, where $B_\text{DIFI}[t]$ is the original DIFI time-series prediction. The solution that minimizes the RSS (Equation \ref{eq:RSS}) will also reduce the RMSE. Thus, we solve for,
\begin{equation}
    b_\text{opt} = \underset{b}{\text{argmin}}\sum_{t}(B_l[t]-( B_\text{DIFI}[t] - b))^2, 
\end{equation} 
where $B_l[t]$ is the ground truth at the local reference station (FRD) and $b_\text{opt}$ is the optimal offset. Analytically, the optimal solution is: 
\begin{equation}
b_\text{opt} = \text{mean}(B_\text{DIFI}[t]-B_l[t]). 
\end{equation}
Thus, we call $B_\text{adj}[t]$ the mean-adjusted DIFI prediction. Figure \ref{fig:difi_sample_pred_mean_adj} demonstrates the improved performance. 


\begin{figure*}
    \centering
    \includegraphics[width=\linewidth]{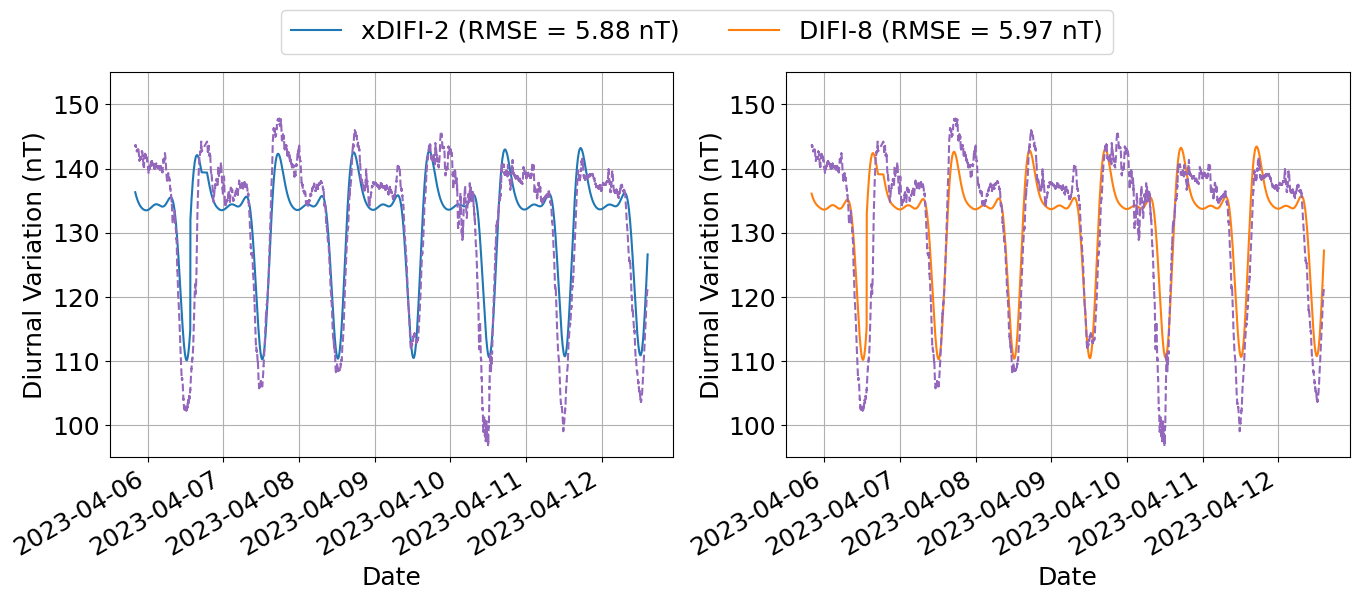}
    \caption{The sample predictions made by DIFI-8 and xDIFI-2, with an adjusted time offset and the mean removed. We observe that the performance improves compared to Figure \ref{fig:difi_sample_pred_mean_adj}. However, ERSM performs better with the Neural-Net and kNN approaches, yielding RMSEs of 3.83 nT and 3.90 nT, respectively.}
\label{fig:difi_lag_sample}
\end{figure*}

Notably, Figure \ref{fig:difi_sample_pred_mean_adj} demonstrates that a time offset hinders DIFI's performance. To consider its performance without the time offset, we computed the optimal time offset (via cross-correlation) and applied that to the DIFI prediction. With the time-offset correction, the results improve in Figure \ref{fig:difi_lag_sample}. However, ERSM still performs better than DIFI, as it predicts higher-frequency components, unlike DIFI, which focuses on the Sq field. 

%% file: src/acknowledgement.tex
DISTRIBUTION A: Public Release: Distribution Unlimited.

This research was supported in part by an appointment to the Department of Defense (DOD) Research Participation Program administered by the Oak Ridge Institute for Science and Education (ORISE) through an interagency agreement between the U.S. Department of Energy (DOE) and the DOD. ORISE is managed by ORAU under DOE contract number DE-SC0014664. All views expressed in this paper are the authors' and do not necessarily reflect the official policies, guidance, or positions of the United States Government, Department of Defense, Department of Energy, Department of the Air Force, United States Air Force, United States Space Force, Air University, or ORAU/ORISE. 


